\documentclass[preprint,number,sort&compress]{elsarticle}

\usepackage{natbib}
\usepackage{appendix}
\usepackage{graphicx}
\usepackage{amsmath}
\usepackage{mathcomp}
\usepackage{MnSymbol}
\usepackage[dvipsnames]{xcolor}
\usepackage{varwidth}
\newcommand{\br}{\mathbf{r}}
\newcommand{\bR}{\mathbf{R}}
\newcommand{\bk}{\mathbf{k}}

\begin{document}

\begin{frontmatter}
\title{Efficient first principles simulation of electron scattering factors for transmission electron microscopy}

\author[vie]{Toma Susi}
\ead{toma.susi@univie.ac.at}
\cortext[cor1]{Corresponding author.}

\author[dtu]{Jacob Madsen}
\author[vie]{Ursula Ludacka}
\author[dtu]{Jens J{\o}rgen Mortensen}
\author[vie,mpi]{Timothy J. Pennycook}
\author[ulm]{Zhongbo Lee}
\author[vie]{Jani Kotakoski}
\author[ulm]{Ute Kaiser}
\author[vie]{Jannik C. Meyer}

\address[vie]{University of Vienna, Faculty of Physics, Boltzmanngasse 5, 1090 Vienna, Austria}
\address[dtu]{Technical University of Denmark, Department of Physics, 2800 Kgs. Lyngby, Denmark}
\address[mpi]{Max Planck Institute for Solid State Research, Stuttgart Center for Electron Microscopy, Heisenbergstra{\ss}e 1,70569 Stuttgart, Germany}
\address[ulm]{Ulm University, Electron Microscopy Group of Materials Sciences, Albert-Einstein-Allee 11, 89081 Ulm, Germany}

\date{\today}

\begin{abstract}
Electron microscopy is a powerful tool for studying the properties of materials down to their atomic structure. In many cases, the quantitative interpretation of images requires simulations based on atomistic structure models. These typically use the independent atom approximation that neglects bonding effects, which may, however, be measurable and of physical interest. Since all electrons and the nuclear cores contribute to the scattering potential, simulations that go beyond this approximation have relied on computationally highly demanding all-electron calculations. Here, we describe a new method to generate \textit{ab initio} electrostatic potentials when describing the core electrons by projector functions. Combined with an interface to quantitative image simulations, this implementation enables an easy and fast means to model electron microscopy images. We compare simulated transmission electron microscopy images and diffraction patterns to experimental data, showing an accuracy equivalent to earlier all-electron calculations at a much lower computational cost.
\end{abstract}

\begin{keyword}
TEM \sep QSTEM \sep DFT \sep 2D materials
\end{keyword}
\end{frontmatter}

\section{Introduction}
Recent decades have seen enormous advances in electron microscopy instrumentation, steadily increasing its power as a central tool for materials science~\cite{Haider1998,Krivanek1999,Haguenau2003a,Nellist04S,Hawkes_aberrationsReview_2009,Krivanek08UM,Urban2009,Erni2010b}. Accurate modelling of electron scattering from solids can be crucial for the interpretation of images and electron diffraction data, and hence, key to obtaining insights to the studied materials. In electron diffraction and phase contrast imaging modes such as electron ptychography and high resolution transmission electron microscopy (HRTEM) the observed contrast is dominated by the phase shifts accumulated by the fast electrons that traverse the sample. These shifts can be derived from the electrostatic potential within the sample\cite{Buseck1992,Spence2009}.

Different approximations for the sample potential exist. In the simplest approximation, a screened Coulomb (SC) potential can be used~\cite{Wentzel1926,Ruhle2003,Shevitski2012}, which has the advantage that the electron scattering factor can be expressed in a simple analytic form. However it neglects any details of the atomic electrons' arrangement into specific orbitals. The independent atom model (IAM) approximates the sample as a superposition of electrostatic potentials previously calculated by first principles for an isolated atom of every element, with several different numerical parameterizations available in the literature~\cite{Doyle1968,Weickenmeier91ACA,Peng96ACA,Kirkland98,Lobato14ACA}. Naturally, this approximation neglects any changes in the electronic charge density that results from interatomic interactions.

For periodic structures, electron and x-ray diffraction patterns are sensitive to the charge transfer in chemical bonds~\cite{Koritsanszky2001,Wu1999,Shibata1999,Zuo1999}. The analysis of such measurements requires a description beyond the IAM, and can provide insights not only into the atomic configuration but also into the electronic structure of a material. The difference between the IAM and a first principles simulation is in many cases large enough to be directly detectable in HRTEM images~\cite{Deng2006,Deng2007}, and can be crucial for the interpretation of small differences in the atomic contrast~\cite{Kurasch11BJoN}. The need for a first-principles based simulation for interpreting the small contrast differences between boron, carbon and nitrogen in experimental HRTEM data was demonstrated earlier by some of the authors of the current work~\cite{Meyer11NM}.

Nearly all previous works have utilized computationally expensive all-electron density functional theory (DFT)~\cite{Exner1998,Mogck2004,Deng2006}, apart from a recent effort to extract the electrostatic potential from a pseudopotential calculation~\cite{Borghartdt17PRL}. However, since that method does not explicitly give the core electron charge needed for TEM simulations, the authors had to resort to a correction scheme~\cite{Wei13PRB}. By contrast, the projector-augmented waves we use here allow the exact (frozen) core electron density to be recovered. Our approach is thus a simple and accurate way to calculate the electrostatic potential in the sample and the subsequent electron scattering factors.

For TEM image simulation, we compare the results of our method to previous experiments and to all-electron DFT based simulations. In addition, by comparing multiple multiple IAM parameterizations, we show that experimental electron diffraction patterns of graphene and hexagonal boron nitride (hBN) can be simulated with a near-perfect match only when bonding effects are taken into account.

\section{Methods}
\subsection{Theory}\label{method}
In the projector-augmented wave (PAW) formalism~\cite{Blochl94PRB}, the total charge density $\rho(\br)$ is a sum of the squared all-electron valence wave functions, the frozen core electron density, and the nuclear charges. For practical calculations, the charge density is divided into a smooth part $\tilde{\rho}(\br)$ plus corrections for each atom $a$: $\rho^a(\br) - \tilde{\rho}^a(\br)$, where the smooth part is given in terms of pseudo wave functions and pseudo core charges. By construction, the multipole moments of $\rho^a(\br) - \tilde{\rho}^a(\br)$ are zero and therefore the electrostatic potential from these correction charges will be non-zero only inside the atomic augmentation spheres. 

This allows us to solve the Poisson equation in two separated steps to obtain the electrostatic potential $v(\br)$. First for the ``pseudo" part,

\begin{equation}
   \nabla^2 \tilde{v}(\br) = -4\pi\tilde{\rho}(\br),
\end{equation}
solved for in all of space on a uniform 3D grid. As second step, the corrections to  $\tilde{v}(\br)$ are added, via

\begin{equation}
   \nabla^2 \Delta v^a(\br) =
   -4 \pi [\rho^a(\br) - \tilde{\rho}^a(\br)],
\end{equation}
which is solved for on a fine radial grid inside the atomic spheres, here only taking the spherical part of the density into account.

As a final approximation, we broaden the nuclear charges by Gaussian functions (width 0.005~\AA) in the total charge density to avoid the corrections diverging as $-Z^a/r$ near the nuclei. A detailed description is given in Appendix~\ref{theory}.

\subsection{Simulation}
To simulate electron microscopy images and diffraction patterns, we use the recently implemented PyQSTEM interface to the Quantitative TEM/STEM Simulations (QSTEM) code~\cite{Koch2002}. Electron scattering is modelled by dividing the simulation cell into slices in the direction perpendicular to the direction of the electron beam, and calculating the propagation of the electron waves from the projected electrostatic potential in each slice to the next. A description of the multislice propagation method can be found in Ref.~\cite{Koch2002}. In the case of the IAM model, a potential is numerically generated based on the positions and atomic species of the modelled material, with parameterizations of Weickenmeier~\cite{Weickenmeier91ACA}, Peng~\cite{Peng96ACA}, Kirkland~\cite{Kirkland98} and Lobato~\cite{Lobato14ACA} available in PyQSTEM in addition to the default QSTEM choice of Rez et al.~\cite{Rez94ACA}. In the \textit{ab initio} approach, the potential is instead derived from the ground state electron and nuclear charge density obtained from DFT (or another first principles simulation method). In the present work, we use the PAW-based code GPAW~\cite{Mortensen05PRB,Enkovaara2010}, which we compare to earlier Wien2k calculations.~\cite{Kurasch11BJoN} Finally, to most directly assess the role of chemical bonding, we parameterized an additional IAM potential based on isolated-atom GPAW calculations.

\subsubsection{PyQSTEM}
PyQSTEM is a Python based interface and extension to the multislice simulation program QSTEM. It was created with the goal of providing a single scripting environment for doing everything related to image simulation, from model building to analysis. This allows simulating large numbers of automatically generated models required for purposes such as statistical analysis, optimization and machine learning. PyQSTEM provides a large degree of flexibility by letting the user supply any custom wave function or potential, and is especially convenient with GPAW.

Python is a well suited interface language due to its prevalence in data science. The numerous extension packages such as numpy~\cite{numpy} and scipy~\cite{scipy} provide direct access to tools for image analysis. The Atomic Simulation Environment (ASE)~\cite{Larsen17JPCM} is used for building atomic models. This is a popular tool in the computational materials community, with modules for defining a wide range of different structures. By using ASE it is easy to integrate results from atomistic simulations into microscopy simulations. The PyQSTEM program and all its dependencies are open source under the GNU general public license~\cite{gnu}, and available on all platforms~\cite{PyQSTEM}.

\subsubsection{TEM simulation with DFT potential}
To simulate TEM images and electron diffraction patterns, we start by building orthorhombic unit cells on the $xy$-plane using ASE, as PyQSTEM assumes propagation along the $z$-direction. We assign a GPAW calculator created in the finite difference (FD) mode where the wave functions are represented on a real space grid.

We then run a DFT calculation with the PBE functional. When this is finished, we can extract the all-electron potential, using the method described in Section~\ref{method} dubbed \texttt{PS2AE} (PSeudo wave to All-Electron wave). The resulting numpy array, \texttt{v}, describes the all-electron potential on a 3D grid in ASE units (eV). The multislice algorithm requires slices of this potential projected along the beam direction 
\begin{align}
v_{proj}^{(i)}(x,y) = \int_{z_i}^{z_i+\Delta z} v(x,y,z) \ dz , 
\end{align}
where $v_{proj}^{(i)}$ is the $i^{th}$ slice and $\Delta z$ is the slice thickness. By also supplying the unit cell, we fix the lateral sampling rate of the TEM simulation.

We then use PyQSTEM in TEM mode; other modes currently supported are STEM and convergent-beam electron diffraction. We set the potential and build a plane wave function at an acceleration voltage of 80 kV. Finally, we run a multislice simulation, propagating the wave function through the potential once.

We can also directly obtain the electron diffraction pattern, which is very useful for quantitative comparison to experiment, as the absolute square of the Fourier transform of the exit wave (a logarithm is easier to visualize, but less useful for the quantification of diffraction intensities). The code that we use is provided in the Supplemental Materials.

\subsection{Experiment}
To compare the simulations with HRTEM images, we refer to published work on hexagonal boron nitride (hBN)~\cite{Meyer11NM}. To further quantify the accuracy, we compare our simulations to electron diffraction measurements of mechanically exfoliated single-layer graphene and single-layer hBN synthesized by chemical vapor deposition~\cite{Caneva2016,Bayer17AN}. The diffraction patterns were recorded on an aberration-corrected FEI Titan 80-300 and on a Philips CM200 microscope (both operated at 80~kV).

\section{Results and discussion}

\subsection{Benchmarking}
\begin{figure}[t!]
\centering
\includegraphics[width=0.75\linewidth]{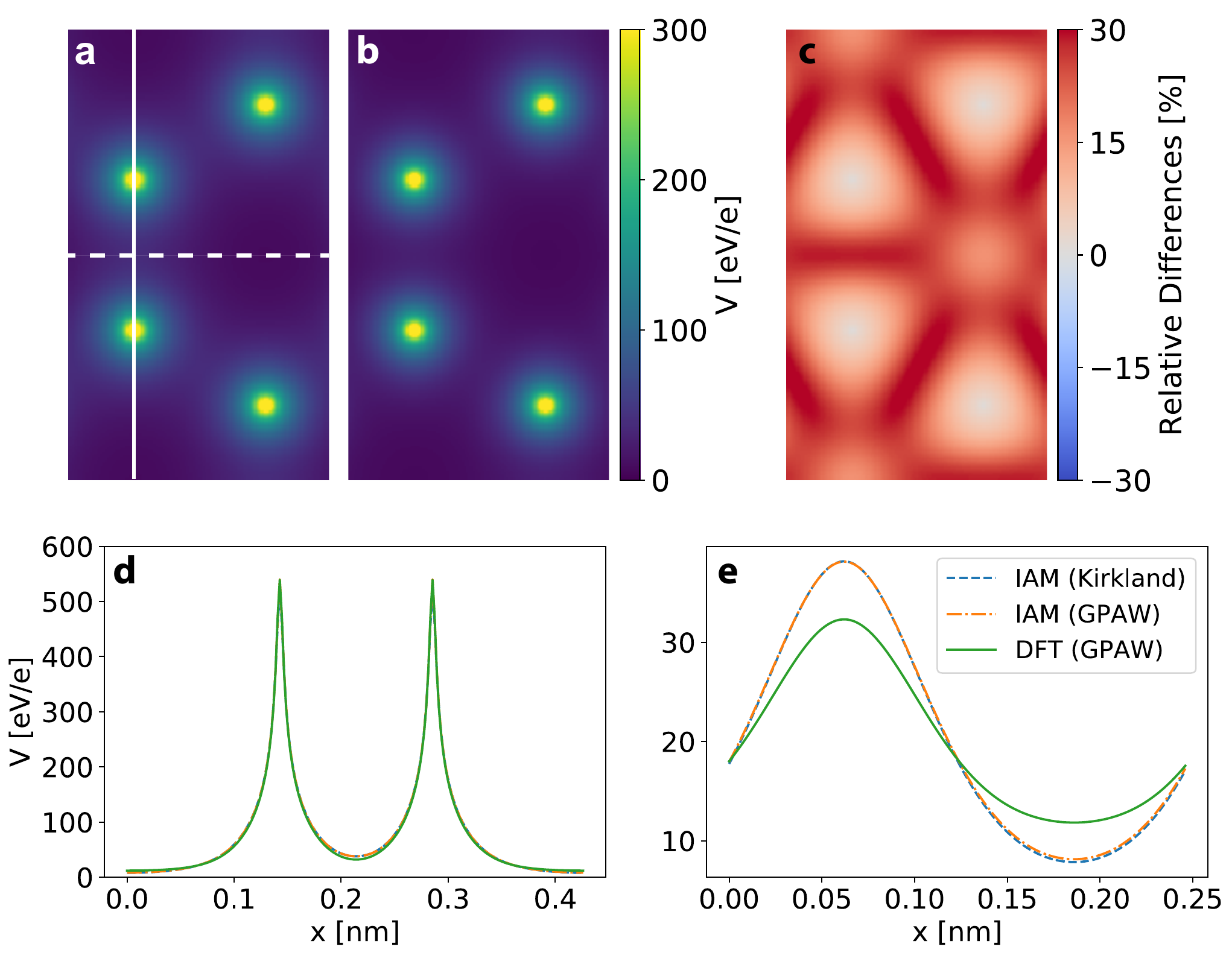}
	\caption{The electrostatic potential of a four-atom unit cell of graphene. a) The DFT-based independent atom model (IAM). b) DFT electrostatic potential calculated with GPAW. c) The relative difference (IAM (GPAW)$-$DFT (GPAW)) / IAM (GPAW). d) Line profiles plotted along the solid line indicated in panel a. e) Line profiles plotted along the dashed line in panel a. (The Kirkland IAM is nearly identical to the GPAW IAM and is thus not shown above.)}
	\label{fig:potential_comparison}
\end{figure}

In Fig.~\ref{fig:potential_comparison} we show the potential of graphene calculated using the Kirkland IAM, our GPAW-based IAM, and via first principles with GPAW as specified above. Since the latter two are based on the same method for calculating the electrostatic potential but one includes chemical bonding, this serves as a direct comparison of its influence. It may seem counter-intuitive that the IAM potential is greater along the C-C bonds (Fig.~\ref{fig:potential_comparison}). However, this is because the main effect of the electron density is to screen the $1/r$ Coulomb potential of the cores~\cite{Wu2004a}. Consequently, bonding concentrates the electron density into the near core regions and between the atoms, reducing the total potential in these regions.

\begin{figure*}[b!]
	\includegraphics[width=1.0\textwidth]{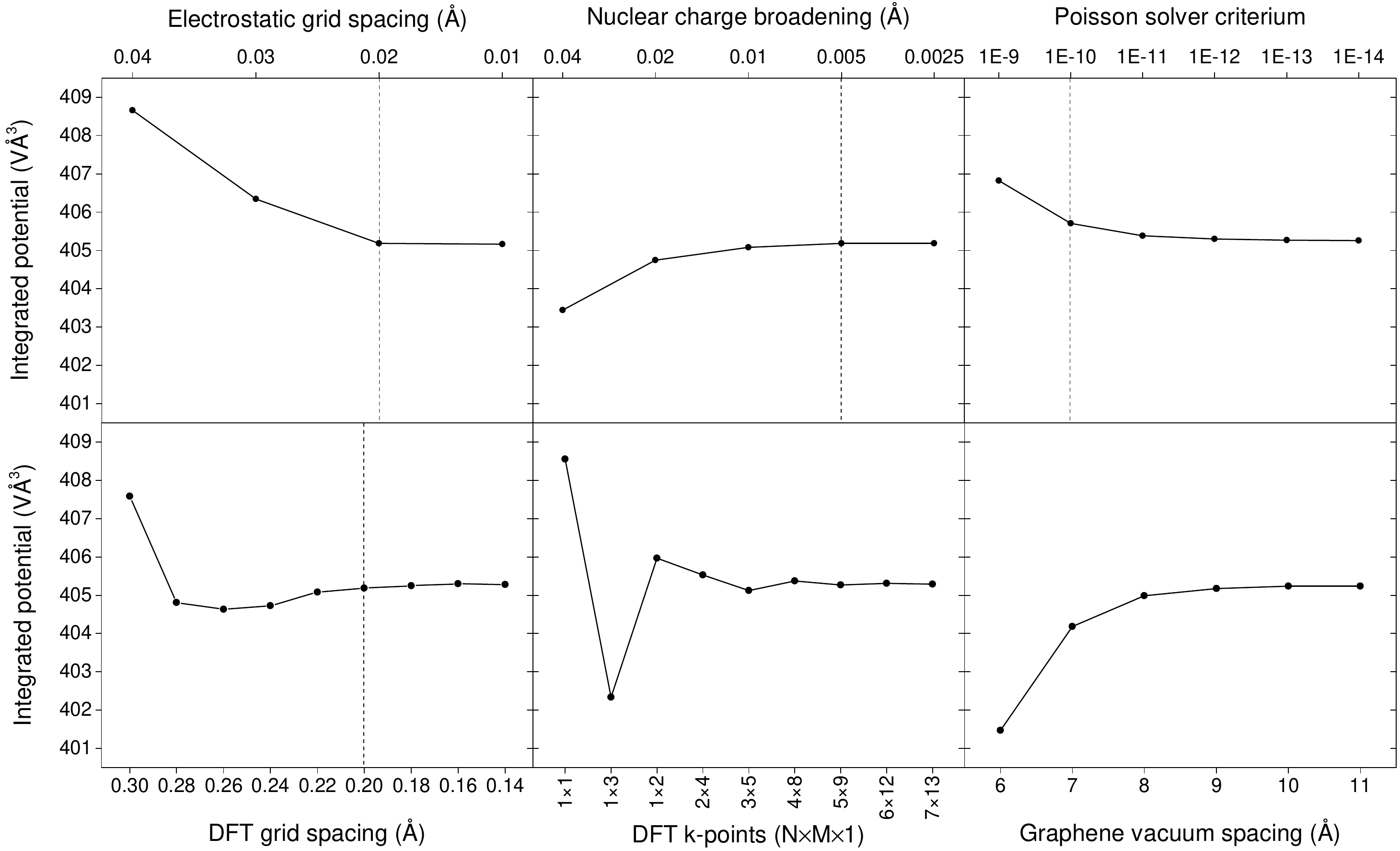}
	\caption{Computational convergence of the integrated electrostatic potential of graphene. The dashed vertical lines indicate GPAW default settings.}
	\label{fig:convergence}
\end{figure*}

We further studied how the DFT parameters affect the integrated electrostatic potential of the four-atom orthogonal unit cell of graphene, calculated using the PBE functional~\cite{Perdew96PRL} (Fig.~\ref{fig:convergence}; LDA yields $\sim$0.5\% higher values). For this measure, we find full convergence with an electrostatic grid spacing\footnote{The electrostatic grid spacing sets the real-space density of the arrays used to describe the electrostatic potential and is important for solving the Poisson equation accurately.} of 0.02~\AA\ and a nuclear charge broadening of 0.005~\AA, a computational grid spacing\footnote{Defines the real-space density of the arrays to describe the electron density numerically, fulfilling a similar convergence role as a plane wave cutoff energy, but is not strictly variational.} of 0.16~\AA\ (the default 0.2~\AA\ is practically converged), a $k$-point mesh finer than 7$\times$13$\times$1, and a Poisson solver convergence criterion\footnote{This criterion is similar to convergence criteria for wave functions and electron density, but instead for solving the Poisson equation within each step of the self-consistency cycle.} of 10$^{-12}$, slightly tighter than the default. For graphene, at least 10 \AA\ of vacuum is required. In the following, we use fully converged parameters.

\subsubsection{Comparison to Wien2k}
\begin{figure}[t!]
	\begin{center}\includegraphics[width=0.75\linewidth]{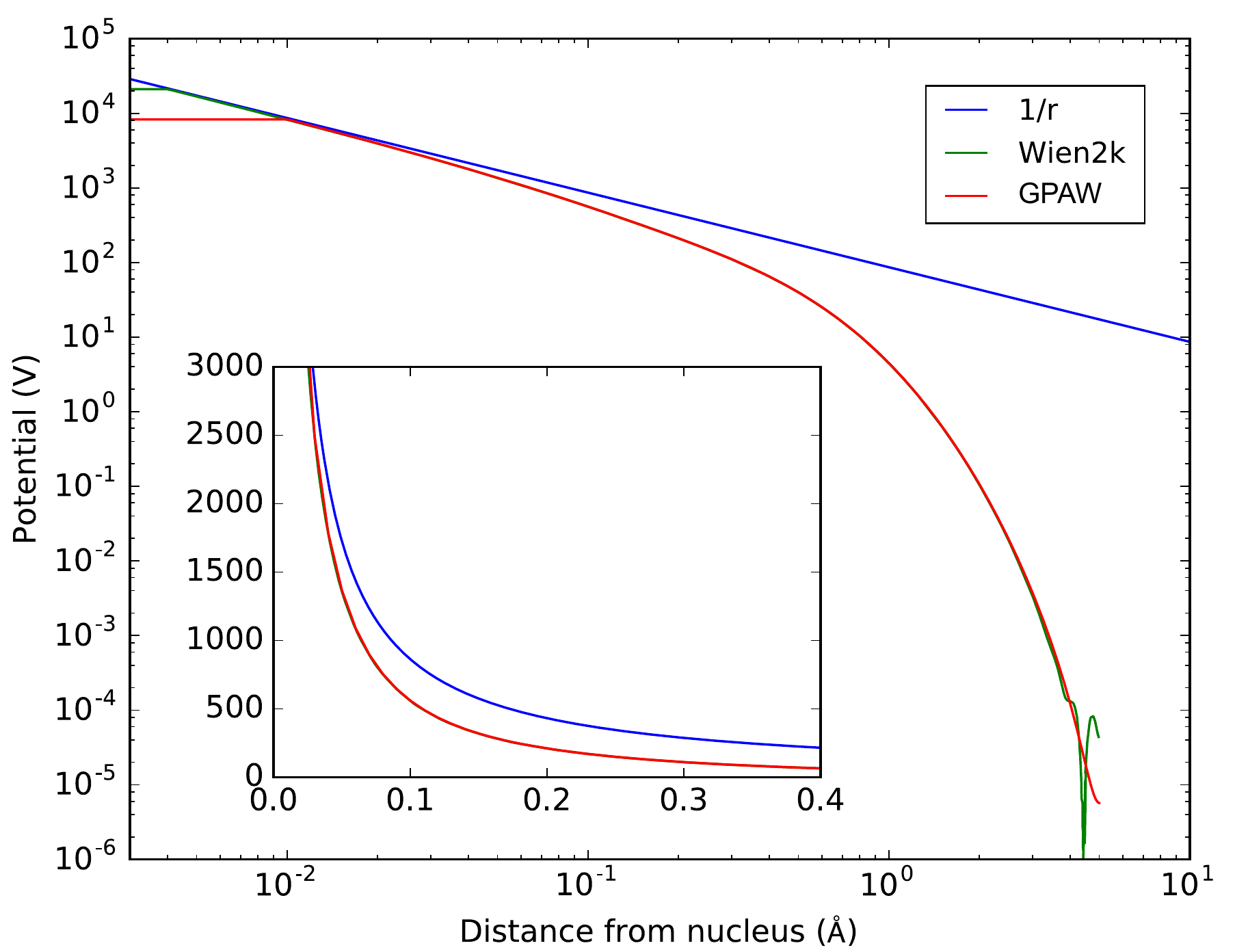}\end{center}
	\caption{Comparison of GPAW electrostatic potential to Wien2k (Ref.~\citenum{Kurasch11BJoN}) and to the pure Coulomb potential of a C nucleus in the center of a 10~\AA\ box.}
	\label{fig:wien2k}
\end{figure}

\begin{figure}[ht!]
	\begin{center}\includegraphics[width=0.65\linewidth]{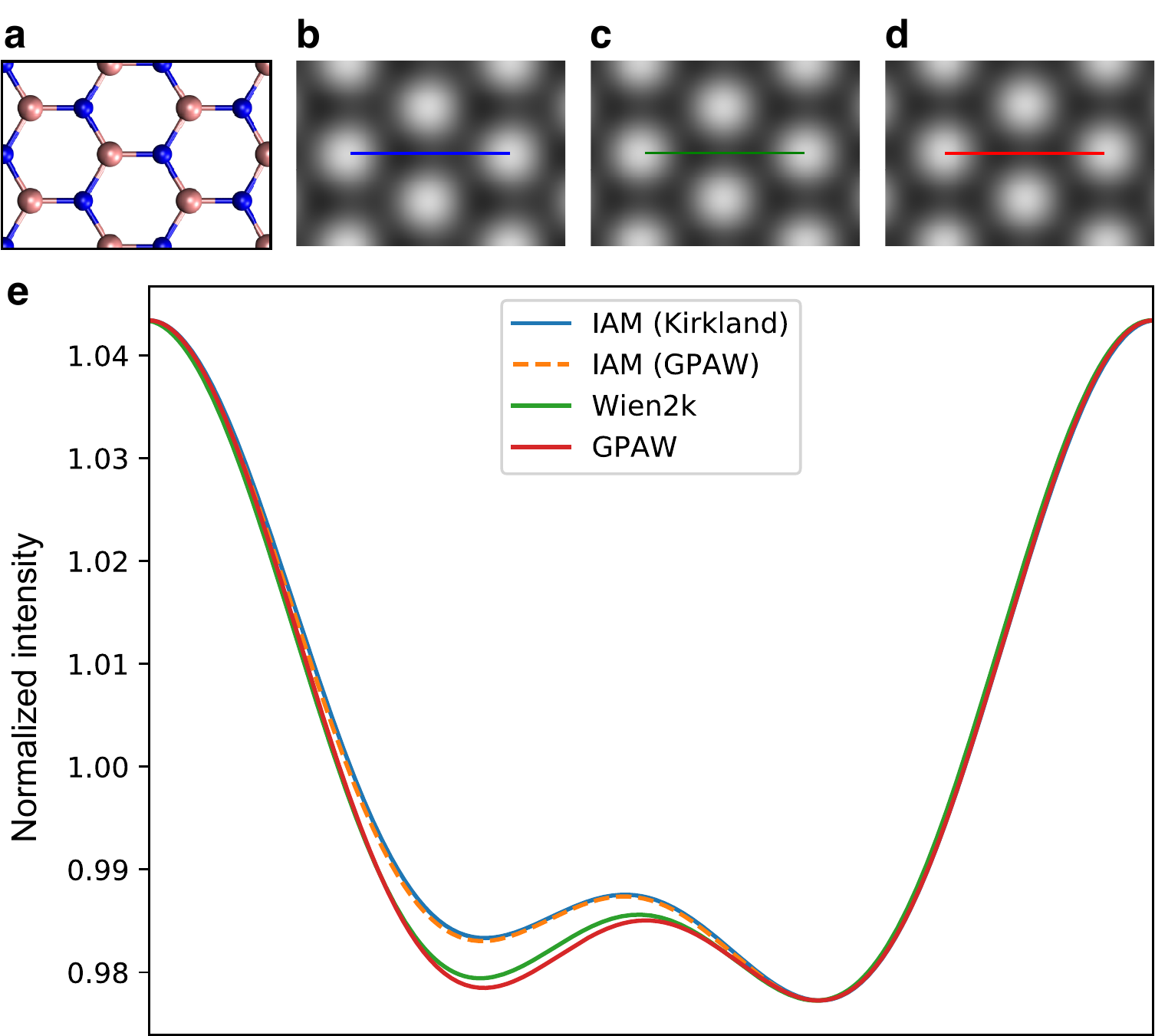}\end{center}
	\caption{Comparison of different simulations for the 80 kV HRTEM image of hBN for a defocus value of $-$9 nm. a) Cropped view of the hBN structure model (boron is pink, nitrogen blue). b) Kirkland independent atom model (IAM). c) Wien2k, as reported in Ref.~\citenum{Meyer11NM}. d) GPAW, this work. e) Profiles over the colored lines in b-d (and the GPAW IAM, which is indistinguishable from the Kirkland one), showing that the GPAW potential results in an even smaller asymmetry over the B and N atoms than previously obtained with Wien2k.}
	\label{fig:hBNtem}
\end{figure}

It is of interest to compare our results to an established all-electron method, such as the Wien2k code (as described in Ref.~\citenum{Kurasch11BJoN}). For the potential near a C nucleus, apart from the slightly different low-distance cutoff (determined by the electrostatic grid spacing, here 0.01 \AA) and minor numerical variation at the periodic cell boundary, the results are identical (Fig.~\ref{fig:wien2k}). In Fig.~\ref{fig:hBNtem} we further compare HRTEM simulations of hBN as reported experimentally and using Wien2k in Ref.~\citenum{Meyer11NM}. Due to a neglect of bonding effects, the IAM predicts a significant asymmetry in the image contrast over the B and N sites. However, an image simulated using the full electrostatic potential derived here correctly predicts a much lesser asymmetry, in a excellent agreement with previous results from Wien2k and with the experiment shown in Ref.~\citenum{Meyer11NM}.

Despite this identical accuracy, our method is significantly faster. Using our converged parameters, a full calculation from a relaxed hBN structure to its electrostatic potential takes only 6 min, compared to 142 min for Wien2k running on the same hardware. An image simulation depends on the number of slices, and only takes some minutes. Furthermore, GPAW scales efficiently to far more cores and larger systems.

\subsubsection{Electron diffraction}
\begin{figure}
\centering
\includegraphics[width=0.65\linewidth]{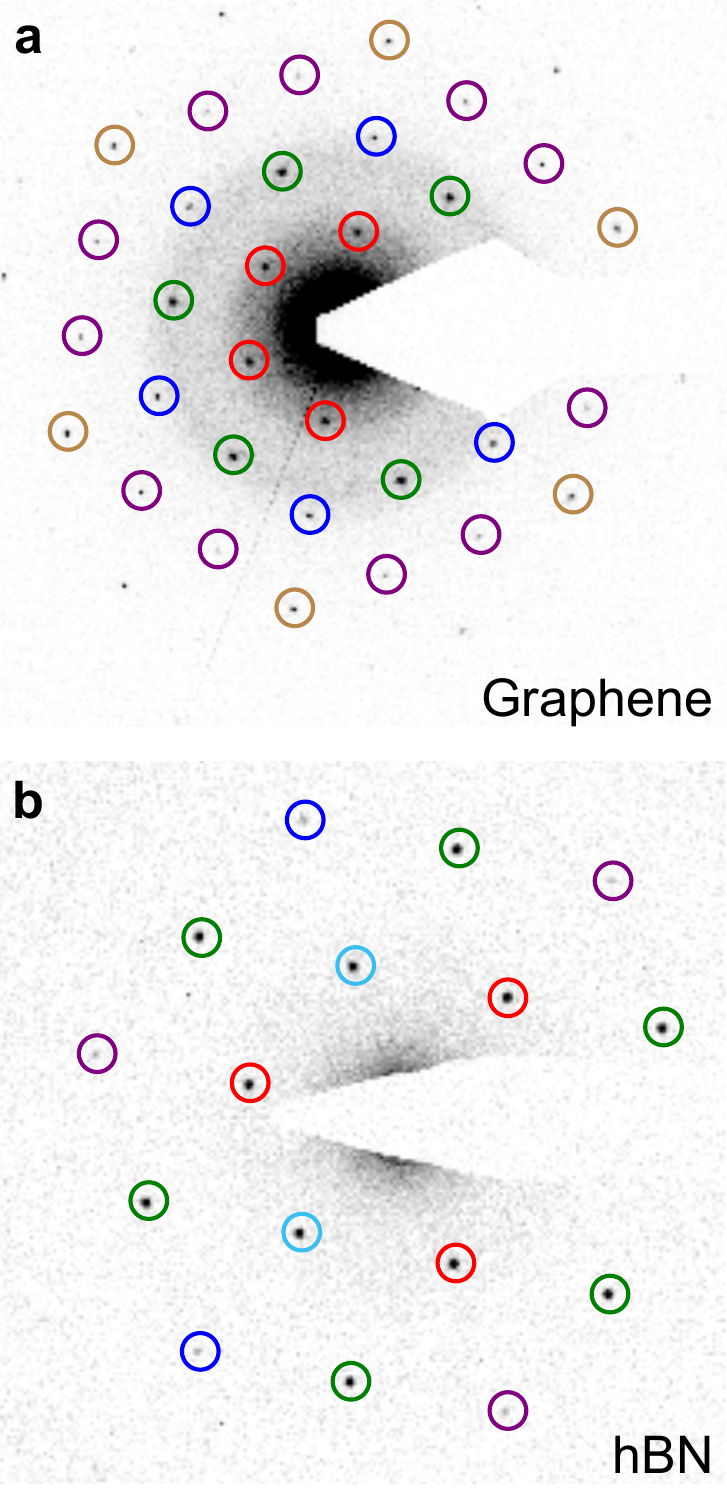}
\caption{(a) Experimental electron diffraction pattern from single-layer graphene. Symmetry-equivalent diffraction peaks are labeled in the same color. Ratios of peak integrated intensities with respect to the red labeled spots are given in the table. (b) Experimental electron diffraction pattern from single-layer hBN. The structure has a lower symmetry than graphene, and hence, inequivalent spots appear in the same diffraction order.\label{fig:diffraction}}
\end{figure}

The graphene diffraction pattern (Fig.~\ref{fig:diffraction}a) exhibits the six-fold symmetry of the lattice. We compare the relative intensities of the first and second set of diffraction peaks, averaging over equivalent peaks and using the innermost ring as reference (intensity = 1). It is important to emphasize that we measure the integrated intensity of each diffraction peak (minus surrounding background) rather than the peak intensity of a line profile, which would be affected by peak broadening.

In high-quality graphene, the ratio of the second- to first-order peaks is very close to 1.0, sometimes even slightly higher (in the pattern in Fig.~\ref{fig:diffraction}, the ratio is 1.04). Highly defective graphene (e.g. graphene oxide), by contrast, shows a much lower ratio~\cite{Gomez-Navarro2010,Pacile2011}, which can be attributed to a static Debye-Waller factor (DWF) resulting from a structure with imperfect periodicity. The IAM predicts a ratio of 0.89, while the first-principles calculation predicts a ratio of 1.10. Considering that any DWF (static from disorder, or dynamic from atomic motion) can only reduce the ratio, the IAM is in clear conflict with our experimental values. However, for the first principles calculation, including the DWF as measured in Ref.~\citenum{Shevitski2012} results in an excellent agreement with experiment.

The single-layer hBN diffraction pattern (Fig.~\ref{fig:diffraction}b) exhibits the expected three-fold symmetry in the innermost ring, a six-fold symmetry in the second ring, and again three-fold symmetry in the third ring (hexagonal symmetry with two inequivalent atoms). We use the weaker inner spot as reference (intensity = 1) and measure (Fig.~\ref{fig:diffraction}) the intensities of the other peaks. Our first principles calculation is in near-perfect agreement for the experimental first, third and fourth intensity ratios, but diverges slightly for the second one (1.16 vs. 1.07). However, including the DWFs again brings the simulated intensity into excellent agreement with experiment.

Tables~\ref{tab:EDgra} and \ref{tab:EDhBN} lists the experimental and simulated diffraction peak intensities. We use DWFs for graphene and for hBN from Ref.~\citenum{Shevitski2012}, given as ratios of the DWF with respect to the reference peak. The lower parts of the tables show the simulated diffraction intensities multiplied with this DWF ratio, which is the set of values that should be compared to the experiment. Multiplying the simulated ratios with the DFWs brings the sum of errors in the first four ratios to within 1\% of experiment for graphene and 2\% for hBN. This remarkably good agreement can only be achieved with a DFT electrostatic potential.

\begin{table}[ht!]
\centering
\caption{Analysis of electron diffraction patterns of graphene from experiment (Fig.~\ref{fig:diffraction}a) and different simulation methods (including IAM parameterizations by Kirkland~\cite{Kirkland98}, Lobato~\cite{Lobato14ACA}, Weickenmeier~\cite{Weickenmeier91ACA},
Peng~\cite{Peng96ACA}, Rez~\cite{Rez94ACA}, and directly from GPAW). {\color{red}$\bigcircle$} = 1} \label{tab:EDgra}
\begin{tabular}{|l c c c c|c}\hline
Graphene & {\color{OliveGreen} $\mathbf{\bigcircle}$} &{\color{blue} $\bigcircle$} &{\color{Fuchsia} $\bigcircle$}&{\color{brown} $\bigcircle$} 
\\\hline
Experiment & 1.03 & 0.16 & 0.05 & 0.12\\
\hline
Wien2k & 1.10 & 0.18 & 0.06 & 0.15 \\
GPAW & 1.11 & 0.18 & 0.06 & 0.15 \\
IAM (GPAW) & 0.99 & 0.16 & 0.07 & 0.18\\
IAM (Kirkland) & 0.98 & 0.16 & 0.06 & 0.17 \\
IAM (Lobato) & 0.98 & 0.16 & 0.06 & 0.17 \\
IAM (Weickenmeier) & 0.93 & 0.15 & 0.07 & 0.18 \\
IAM (Peng) & 0.92 & 0.14 & 0.06 & 0.14 \\
IAM (Rez) & 0.89 & 0.14 & 0.05 & 0.12 \\
\hline
DWF ratio & 0.93 & 0.89 & 0.79 & 0.74 \\
\hline
Wien2k$\times$DWF & 1.02 & 0.14 & 0.05 & 0.11 \\
GPAW$\times$DWF & 1.03 & 0.14 & 0.05 & 0.11 \\
IAM (Kirkland)$\times$DWF & 0.91 & 0.14 & 0.05 & 0.13 \\
\hline
\end{tabular}
\end{table}

\begin{table}[h!]
\centering
\caption{Analysis of electron diffraction patterns of hBN from experiment (Fig.~\ref{fig:diffraction}b) and different simulation methods. {\color{red}$\bigcircle$} = 1} \label{tab:EDhBN}
\begin{tabular}{|l c c c c|c}\hline
hBN &{\color{cyan} $\bigcircle$} &{\color{OliveGreen} $\bigcircle$} &{\color{blue} $\bigcircle$} &{\color{Fuchsia} $\bigcircle$} \\\hline
Experiment & 1.05 & 1.07 & 0.19 & 0.19 \\
\hline
Wien2k & 1.05 & 1.17 & 0.16 & 0.17 \\
GPAW & 1.06 & 1.16 & 0.16 & 0.17 \\
IAM (GPAW) & 1.07 & 0.95 & 0.17 & 0.16 \\
IAM (Kirkland) & 1.07 & 0.96 & 0.17 & 0.16 \\
IAM (Lobato) & 1.07 & 0.96 & 0.17 & 0.16 \\
IAM (Peng) & 1.06 & 0.94 & 0.16 & 0.16 \\
IAM (Weickenmeier) & 1.07 & 0.94 & 0.17 & 0.16 \\
IAM (Rez) & 1.07 & 0.95 & 0.13 & 0.13\\
\hline
DWF ratio & 1 & 0.93 & 0.89 & 0.89 \\
\hline
Wien2k$\times$DWF & 1.05 & 1.09 & 0.15 & 0.16 \\
GPAW$\times$DWF & 1.06 & 1.08 & 0.15 & 0.16 \\
IAM (Kirkland)$\times$DWF & 1.07 & 0.89 & 0.15 & 0.14\\
\hline
\end{tabular}
\end{table}

\section{Conclusions}
Efficient simulation of the full electrostatic potential of materials is becoming ever more important with the development of better instrumentation. Approximate models, though feasible for large systems and sufficient for routine simulations, do not capture bonding effects that can in some cases be directly measured, nor are they sufficient for electron holography. We have shown how the electrostatic potential derived from frozen core projector-augmented wave density functional theory gives a description of electron scattering equal to previously available and significantly more demanding methods such as Wien2k, and in excellent agreement with experiment on graphene and hexagonal boron nitride.

Due to its computational efficiency, our approach opens the way for the treatment of large systems with defects, such as impurities or grain boundaries, and is not limited to two-dimensional specimens. Although we have chosen to concentrate on high-resolution transmission electron microscopy due to its greater sensitivity to bonding effects, scanning TEM images can equally well be simulated. The current implementation provides a convenient computational workflow starting from a structure model all the way to high-quality images or diffraction patterns in one simple script.

\section*{Acknowledgments}
We thank Bernhard C. Bayer for providing the experimental hBN diffraction pattern. T.S. acknowledges the Austrian Science Fund (FWF) for funding via project P 28322-N36, J.C.M. via project P25721-N20, and J.K. via project I 3181-N36. U.L. and J.K were supported by the Wiener Wissenschafts\mbox{-,} Forschungs- und Technologiefonds (WWTF) via project MA14-009. T.J.P acknowledges funding from European Union's Horizon 2020 Research and Innovation Programme under the Marie Sk\l odowska-Curie grant agreement no.~655760--DIGIPHASE. Z.L. and U.K. acknowledge funding from the German Research Foundation (DFG) and the Ministry of Science, Research and the Arts (MWK) of Baden-W\"urttemberg in the frame of the Sub-\AA ngstrom Low-Voltage Electron microscopy (SALVE) project. J.M. acknowledges funding through grant 1335-00027B from the
Danish Council for Independent Research.

\begin{appendix}
\section{Electrostatic potential}\label{theory}
In the projector-augmented wave (PAW) formalism~\cite{Blochl94PRB}, the total charge density (for convenience, we count electrons as positive and protons as negative charge) can be written as

\begin{equation}
    \rho(\mathbf{r}) =
    2 \sum_{n\bk} f_{n\bk}
    |\psi_{n\bk}(\br)|^2 +
    \sum_a n_c^a(|\br-\bR^a|) -
    \sum_a Z^a \delta(\br - \bR^a),
\end{equation}
where $\psi_{n\bk}(\br)$ are the all-electron valence wave functions explicitly included in the calculation ($n$ is the band index and $\mathbf{k}$ is the crystal momentum) and $f_{n\bk}$ are occupation numbers. For atom number $a$, $n_c^a(r)$ is the frozen core electron density, $\bR^a$ is the position and $Z^a$ is the atomic number. 
For practical calculations, the charge density is divided into a smooth part plus corrections for each atom:

\begin{equation}
  \rho(\br) = \tilde{\rho}(\br) +
  \sum_a [\rho^a(\br - \bR^a) -
          \tilde{\rho}^a(\br - \bR^a)],
\end{equation}
where the smooth part is given in terms of pseudo wave functions $\tilde{\psi}_{n\bk}(\br)$, pseudo core charges
$\tilde{n}_c^a(r)$, expansion coefficients $Q_{\ell m}^a$ (to be defined below) and localized shape functions that in the GPAW code\cite{Mortensen05PRB,Enkovaara2010} have been chosen to be Gaussian functions $\tilde{g}_{\ell m}^a(\br)\propto r^\ell\exp(-\alpha^a r^2)Y_{\ell
m}(\hat{\br})$:

\begin{equation}
    \tilde\rho(\br) =
    2 \sum_{n\bk} f_{n\bk}
    |\tilde{\psi}_{n\bk}(\br)|^2 +
    \sum_a \tilde{n}_c^a(|\br-\bR^a|) +
    \sum_a \sum_{\ell m}
    Q_{\ell m}^a \tilde{g}_{\ell m}^a(\br-\bR^a).
\end{equation}
where $\ell$ and $m$ are the azimuthal and magnetic quantum numbers, $Y_{\ell m}(\hat{\br})$ are the spherical harmonics, and the atom-dependent decay factor $\alpha$ is chosen such that the charges are localized within the augmentation sphere.

The corrections look similar to the definitions of $\rho(\br)$ and $\tilde\rho(\br)$ except that we now expand the wave functions in all-electron and pseudo partial waves $\phi_i^a$ and $\tilde{\phi}_i^a$:

\begin{equation}
    \rho^a(\br) =
    \sum_{ij} D_{ij}^a
    \phi_i^a(\br)\phi_j^a(\br) +
    n_c^a(r) -
    Z^a \delta(\br),
\label{rho}
\end{equation}
\begin{equation}
    \tilde\rho^a(\br) =
    \sum_{ij} D_{ij}^a
    \tilde{\phi}_i^a(\br)\tilde{\phi}_j^a(\br) +
    \tilde{n}_c^a(r) +
    \sum_{\ell m}
    Q_{\ell m}^a \tilde{g}_{\ell m}^a(\br).
\end{equation}
The atomic density matrix $D_{ij}^a$ is evaluated from projections of the pseudo wave functions onto smooth PAW projector functions $\tilde p_i^a(\br)$ localized inside the each atomic augmentation sphere:

\begin{equation}
  D_{ij}^a =
  2 \sum_{n\bk}
  \langle \tilde{\psi}_{\bk n} | \tilde{p}_i^a \rangle
   f_{n\mathbf{k}}
  \langle \tilde{p}_j^a | \tilde{\psi}_{\bk n} \rangle.
\end{equation}
The coefficients $Q_{\ell m}^a$ are chosen so that all multipole moments of $\rho^a(\br) - \tilde{\rho}^a(\br)$ are zero and therefore the electrostatic potential from these correction charges will be non-zero only inside the atomic augmentation spheres. 

This allows us to solve the Poisson equation in two separated steps, first for the pseudo part:

\begin{equation}
   \nabla^2 \tilde{v}(\br) = -4\pi\tilde{\rho}(\br),
\end{equation}
solved for in all of space on a uniform 3D grid. 

In the second step, corrections are added to $\tilde{v}(\br)$:

\begin{equation}
   \nabla^2 \Delta v^a(\br) =
   -4 \pi [\rho^a(\br) - \tilde{\rho}^a(\br)],
\end{equation}
solved for on a fine radial grid inside the atomic spheres only taking the spherical part of the density into account. As a final approximation, we replace $\delta(\br)$ in eq.~(\ref{rho}) by $(r_c^2\pi)^{-3/2}e^{-(r/r_c)^2}$ (with $r_c=0.005$~\AA) to avoid the corrections diverging as $-Z^a/r$ near the nuclei.

\end{appendix}


%\bibliographystyle{elsarticle-num}
%\bibliography{References_BibDesk.bib,library.bib}

\end{document}